\documentclass[fleqn,usenatbib]{mnras}

\usepackage{graphicx}	
\usepackage{amsmath}	
\usepackage{amssymb}	
\usepackage{multicol}        
\usepackage{multirow}        
\usepackage{bm}		
\usepackage{pdflscape}	

\newcommand{\referee}[1]{#1}
\renewcommand{\referee}[1]{#1}
\newcommand{\updated}[1]{#1}
\renewcommand{\updated}[1]{#1}
\newcommand{\editor}[1]{#1}
\renewcommand{\editor}[1]{#1}

\usepackage[T1]{fontenc}
\usepackage{ae,aecompl}

\usepackage{txfonts}
\let\la=\lesssim

\title[Resolving the blazar CGRaBS J0809+5341]{Resolving the blazar CGRaBS J0809+5341 in the presence of telescope systematics}

\author[Iniyan Natarajan et al.]{Iniyan Natarajan$^{1}$\thanks{Contact e-mail: \href{mailto:iniyan@ast.uct.ac.za}{iniyan@ast.uct.ac.za}},$\ $
Zsolt Paragi$^{2}$,$\ $
Jonathan Zwart$^{1,\, 3}$,$\ $
Simon Perkins$^{4}$,
\newauthor
Oleg Smirnov$^{5,\, 4}$,$\ $and$\, $
Kurt van der Heyden$^{1}$
\\
$^{1}$Astrophysics, Cosmology and Gravity Centre (ACGC), Department of Astronomy, University of Cape Town, Private Bag X3, Rondebosch 7701, South Africa\\
$^{2}$Joint Institute for VLBI ERIC, Postbus 2, 7990 AA Dwingeloo, the Netherlands\\
$^{3}$Department of Physics and Astronomy, University of the Western Cape, Private Bag X17, Bellville 7535, South Africa\\
$^{4}$SKA South Africa, 4th Floor, The Park, Park Road, Pinelands, 7405, South Africa\\
$^{5}$Department of Physics and Electronics, Rhodes University, PO Box 94, Grahamstown 6140, South Africa\\
}

\date{Last updated 2016 July 13; in original form 2016 July 13}

\pubyear{2016}

\begin{document}
\label{firstpage}
\pagerange{\pageref{firstpage}--\pageref{lastpage}}
\maketitle

\begin{abstract}
We analyse Very Long Baseline Interferometry (VLBI) observations of the blazar CGRaBS J0809+5341 using Bayesian inference methods.
The observation was carried out at 5 GHz using 8 telescopes that form part of the European VLBI Network. Imaging and
deconvolution using traditional methods imply that the blazar is unresolved. To search for source structure beyond the diffraction limit, we 
perform Bayesian 
model selection between three source models (point, elliptical Gaussian, and circular Gaussian).
Our modelling jointly accounts for 
antenna-dependent gains and system equivalent flux densities.
We obtain posterior distributions for the various source and instrumental parameters along with the corresponding uncertainties and correlations between 
them.
We find that there is very strong evidence (>$10^9$:1) in favour of elliptical Gaussian structure and using this model derive the 
apparent brightness temperature distribution of the blazar, accounting for uncertainties in the shape estimates. 
To test the 
integrity of our method, we also perform model selection on synthetic observations and use this to develop a Bayesian criterion for the minimum 
resolvable source size and consequently the maximum measurable brightness temperature for a given interferometer, 
dependent on the signal-to-noise ratio (SNR) of the data incorporating the aforementioned systematics.
We find that calibration errors play an increasingly important role in determining the over-resolution limit for SNR$\gg$100.
We show that it is possible to exploit the resolving power of future VLBI arrays down to about 5 per cent of the size of the naturally-weighted 
restoring beam, if the gain calibration is precise to 1 per cent or less.
\end{abstract}

\begin{keywords}
methods: data analysis -- methods: statistical -- techniques: high angular resolution -- techniques: interferometric -- quasars: individual: CGRaBS J0809+5341
\end{keywords}



\section{Introduction}

The commonly used measure for the angular resolving power of an optical system is the \textit{Rayleigh criterion} \citep{rayleigh1879}, which, 
in its simplest form, states that the minimum resolvable angle is proportional to the quantity $\lambda/d$, where $\lambda$ is the 
wavelength of observation and $d$ is the diameter of the aperture. 
For a given wavelength, the angular resolution can be improved by increasing the aperture size. 
The technique of \textit{aperture synthesis} \citep{ryle1955}, in which multiple telescopes are operated together as a single instrument known as
an \textit{interferometer}, is employed \referee{in radio, optical, and infra-red astronomy} to synthesise a large, 
partially-filled aperture in order to obtain high angular resolution.
This is achieved by computing the \textit{cross-correlation} of the radiation fields received at two different locations, also known as the 
\textit{mutual coherence function}. In radio interferometry, the measured coherence function is a complex-valued 
\textit{visibility}\footnote{When the measurements
are made using two mutually-orthogonal polarisation feeds that record the entire polarisation state of the radio wave, we obtain a \textit{source coherency} or
\textit{visibility matrix} \citep[e.g.,][]{oms2011}. See also section \ref{subsec:rimeandsw}.}, and 
is related to the sky brightness distribution via a Fourier transform operation,
\referee{under some simplifying assumptions such as the field-of-view being small enough \citep[e.g.,][]{Tools2009}.}
By convention,
this Fourier domain in which the visibilities are measured is known as the \textit{visibility} domain or the \textit{uv}-domain.
Any two interferometer antennas (or \textit{stations}) may be said to form a \textit{baseline}, and the length of the longest baseline 
for a given interferometer array configuration becomes the diameter $d$ of the aperture.
Thus, angular resolution of less than a milli-arcsecond (mas) can be achieved using Very Long Baseline Interferometry (VLBI) techniques in 
which data from radio telescopes separated by thousands of kilometres operating independently are synchronised post hoc using atomic clocks 
\citep[e.g.,][]{TMS2001,enno2008}. 

The theoretical limit for the resolution of an interferometer can be obtained using Fourier optics which characterises the resolving power 
of an optical system in terms of its \textit{spatial bandwidth} \citep{goodman1968}.
Diffraction effects limit the maximum spatial frequency, $f_{\mathrm{max}}$, transmitted by the optical system
and the resolving power may be reformulated in terms of this frequency as $R = \pi/f_{max}$, where $R$ is the \textit{Nyquist distance} 
\citep{berterodemol1996}.
It is possible to obtain information about the spatial frequencies that lie outside $f_{max}$ (i.e., beyond the \textit{diffraction limit}) 
by incorporating 
\textit{a priori} knowledge about the brightness distribution of the source. The resolving power then depends on the precision of the 
measuring instrument \citep[e.g.,][]{harris1964,difrancia1969}. 
Introducing prior information to help solve a problem falls under the domain of inverse problem theory, which deals with the question of 
obtaining causes from results \citep{tarantola2005}. Most scientific questions are of this kind. More than one distribution of the relevant
parameter values could have given rise to the same result\footnote{An \textit{ill-posed} problem for which there is no unique solution.} 
and hence it becomes necessary to impose additional constraints in order to obtain the most sensible explanation for the cause. 
This process is known as \textit{regularisation} and is best understood in terms of probability theory. Under the Bayesian probability 
formalism, \textit{a priori} information about a model is introduced in the form of prior distributions on its parameters. 

Techniques for over-resolution or \textit{super-resolution} have been widely used in radio interferometry, and especially in VLBI, to characterise 
partially resolved compact sources on milli-arcsecond scales. 
\citet{lobanov2005} derives the resolution limits for specific brightness-distribution templates for astronomical
sources and \citet{martividal2012} extend this to the general case of super-resolution with interferometers.
\referee{A discussion of model-fitting in the \textit{uv}-domain for specific source profiles using real and synthetic data using the 
\textsc{uvmultifit} package is provided in \citet{uvmultifit2014}. An alternative which achieves over-resolution by \textit{sparse} modelling
\footnote{\referee{\textit{Sparsity} is the ratio between the number of non-zero elements in a matrix and the total number of elements; a sparse matrix
is one whose sparsity is high.}} in the sky or \textit{image} domain in the presence of Gaussian noise generated with the same standard deviation on all baselines 
is explored in \citet{honma2014}.}
While these works explore the theoretical constraints of super-resolution, the application of these methods to
real observations made with a specific interferometer configuration \updated{in the presence of instrumental uncertainties, will not provide 
us with a knowledge of how they correlate with the source parameters and limit the resolution}.
This issue will likely be relevant for snapshot VLBI observations using the Square Kilometre Array (SKA) 
\citep{paragi2015} as well.

\citet{biro2015} present an overview of the history and an implementation of Bayesian inference methods for simultaneous estimation of the sky 
and instrumental parameters through the analysis of visibilities.
This work is organised as follows:
In section \ref{sec:theory}, we explore the theoretical background for performing Bayesian analysis in the \textit{uv}-domain and modelling visibilities
using the measurement equation formalism.
In section \ref{sec:dataanalysis}, we apply this approach to the European VLBI Network (EVN) observations of a flaring blazar CGRaBS J0809+5341 
(henceforth, J0809+5341).
The source is bright, and was \updated{observed for a relatively short time with a limited \textit{uv}-coverage, to study possible structural 
changes associated with the flare \citep{antao2016}.} The situation is complicated by the fact that the station providing the long baselines
 in the array has no short spacings comparable to those provided by the rest of the stations, \referee{making its gain calibration difficult
\citep{martividal2012}.}
Our aim is to explore the resolving power 
of this interferometer and to better understand the uncertainties in the estimated source model parameters 
and how the station calibration errors affect these estimates.
Section \ref{sec:bayesblazar} presents our methodology in detail and section \ref{sec:results}, our results. 
The same data were analysed using traditional methods by \citet{antao2016}.
In section \ref{sec:martividal}, we compare the minimum resolvable source sizes we obtain for this interferometer using the Bayesian approach on 
synthetic observations of compact sources with those predicted by \citet{martividal2012}, in the presence of station-dependent gains and 
baseline-dependent noise terms. 

\section{Theoretical Background}
\label{sec:theory}
\subsection{Bayesian Inference}
\label{subsec:bayes}
Statistical inference is the process of drawing conclusions about scientific propositions from data. A mathematically consistent way of
doing this is by using Bayes' theorem to update our beliefs about propositions as more information becomes available \citep{jaynes2003}. 
In general, given two propositions $A$ and $B$, and relevant prior information $I$, Bayes' theorem states that
\begin{equation}
\label{eq:bayes}
\mathcal{P}(A|B,I) = \frac{\mathcal{P}(A|I)\, \mathcal{P}(B|A,I)}{\mathcal{P}(B|I)}\quad,
\end{equation}
where $\mathcal{P}(X|I)$ denotes the conditional probability of proposition $X$ given that information $I$ is true and `,' denotes the conjunction 
\textit{and}.

Bayesian inference can be performed at two levels \citep{mackay2003}. At the first level (\textit{parameter estimation}), we assume that a model (or 
hypothesis) is true and fit its parameters to the data. If $\Theta$ denotes the set of parameters associated with the model $H$, then, given 
data $D$, Bayes' theorem may be rewritten as
\begin{equation}
\label{eq:bayesparest}
\mathcal{P}(\Theta|D,H) = \frac{\mathcal{P}(\Theta|H)\, \mathcal{P}(D|\Theta,H)}{\mathcal{P}(D|H)}\quad.
\end{equation}
$\mathcal{P}(\Theta|H) \equiv \Pi(\Theta)$ is called the \textit{prior} probability distribution, which encodes our beliefs about the parameters prior to the 
analysis of the data\footnote{The data may be captured before the prior probability distribution is drawn up. This is perfectly fine, since the prior occurs 
\textit{logically} before the posterior in the chain of reasoning, not \textit{temporally}.}. 
$\mathcal{P}(\Theta|D,H)$ 
is the \textit{posterior} probability distribution of the parameters which describes how the data $D$ modify our initial 
beliefs. 
$\mathcal{P}(D|\Theta,H) \equiv \mathcal{L}(\Theta|D,H)$ is called the \textit{likelihood}, which is a function of the parameters given the 
data.  
The mathematical form of the likelihood function reflects how the uncertainties in the measurements of the data are distributed 
\citep{trotta2008}.

The denominator in equation (\ref{eq:bayesparest}), $\mathcal{P}(D|H) \equiv \mathcal{Z}$, is called the \textit{evidence} (or the \textit{marginal likelihood}) and is 
obtained by integrating the numerator over $\Pi(\Theta)$:
\begin{equation}
\label{eq:evidence}
\mathcal{Z} = \int\, \Pi(\Theta)\mathcal{L}(\Theta|D,H)\quad \mathrm{d}^N\Theta\quad,
\end{equation}
where $N$ is the length of $\Theta$. This process of \textit{marginalisation} allows us to
obtain the probabilities of a subset of the parameters without reference to the parameters with respect to which the integration is carried 
out. When the integration is performed over all of $\Theta$, we get the evidence. 
This quantity may safely be ignored during parameter estimation (unless one intends to perform the second level of inference) since it does not 
affect the location or the shape of the posterior distribution in the parameter space and serves only as a normalisation constant.

The second level of inference (\textit{model selection}) is determining
the relative probabilities of alternative hypotheses given the data.
Given hypothesis $H$, and a prior belief in the validity of $H$ given by $\mathcal{P}(H|I)$, 
the model 
posterior probability may be computed using the evidence obtained from parameter estimation as 
\begin{equation}
\label{eq:bayesmodelsel}
\mathcal{P}(H|D,I) \propto \mathcal{P}(D|H,I)\, \mathcal{P}(H|I)\quad.
\end{equation}
The evidence is a quantitative measure of how much the data favour one model over another. It incorporates 
\textit{Occam's razor}\footnote{Occam's razor states that, among competing hypotheses, the one that makes the fewest assumptions should be favoured.} 
automatically: unless a complicated model with more parameters (a high-dimensional parameter space) is significantly better at explaining the data \referee{(i.e., has a higher likelihood)}, 
its evidence will be smaller than that of a simpler model with fewer parameters that can explain the data equally well. 

Given two models $H_1$ and $H_2$, we may define a model selection ratio between the posteriors of the two models as
\begin{equation}
\label{eq:modelselrat}
\frac{\mathcal{P}(H_1|D,I)}{\mathcal{P}(H_2|D,I)} = \frac{\mathcal{Z}_1}{\mathcal{Z}_2}\, \frac{\mathcal{P}(H_1|I)}{\mathcal{P}(H_2|I)} = B_{12}\, \frac{\mathcal{P}(H_1|I)}{\mathcal{P}(H_2|I)}\quad,
\end{equation}
where $\mathcal{P}(H_1|I)/\mathcal{P}(H_2|I)$ is the ratio of the prior probability distributions of the two models which may often be set to unity, 
indicating that there is no prior preference for one model over the other. The ratio of the evidences, $B_{12}$, is known as the 
\textit{Bayes factor} \citep{jeffreys1961}. The higher this factor is, the more is $H_1$ preferred over $H_2$. It is useful to consider the 
natural logarithm of the Bayes factor:
\begin{equation}
\label{eq:logbayesfactor}
\mathrm{ln}(B_{12}) = \mathrm{ln}(\mathcal{Z}_1) - \mathrm{ln}(\mathcal{Z}_2)\quad.
\end{equation}
We use twice this value as a measure of how strongly a model is preferred over another \citep{kassraftery1995} (Table \ref{tab:evidences}).
\begin{table}
\caption{Criteria for model selection. $B_{12}$ denotes the ratio of the evidences between hypotheses $H_1$ and $H_2$ \citep{kassraftery1995}.}
\begin{center}
\begin{tabular}{ c | c | c }
\hline
$2\, \mathrm{ln} (B_{12})$ & $B_{12}$ & Evidence against $H_2$ \\
\hline
$0$ to $2$ & $1$ to $3$ & Not worth more than a mention \\
$2$ to $6$ & $3$ to $20$ & Positive \\
$6$ to $10$ & $20$ to $150$ & Strong \\
$>10$ & $>150$ & Very strong \\
\hline
\end{tabular}
\end{center}
\label{tab:evidences}
\end{table}

Frequently, the distributions involved are analytically intractable. Numerical techniques such as MCMC (Markov Chain Monte Carlo) 
\citep{metropolis, hastings} are used
to sample a multi-dimensional parameter space, resulting in the estimation of the joint posterior distribution of the parameters, but they 
are inefficient when required to compute the Bayesian evidence, since it is a multi-dimensional integral. 
We therefore use the M\textsc{ulti}N\textsc{est}\footnote{https://ccpforge.cse.rl.ac.uk/gf/project/multinest} algorithm \citep{multinest1, multinest2, multinest3}, based on the nested 
sampling method originally proposed by \citet{skilling2004}, 
a Monte Carlo method efficient at sampling posteriors with multiple modes and/or curving degeneracies in low-dimensional parameter spaces
($\lesssim30$). M\textsc{ulti}N\textsc{est} also computes 
the evidence at a fraction of the cost one would incur with the standard MCMC techniques \citep{pymultinest2014}. 

\subsection{Bayesian Analysis of Visibilities}
\label{subsec:statsanalysis}
Estimating the true brightness distribution of the sky from radio data falls under the class of \textit{inverse problems}, which aim to 
determine the underlying phenomena (the causes) from the observed data (the results) \citep{parker1977}.
The technique of \textit{model-fitting} is generally suited to handling such problems. 
It is most useful when the source brightness distribution and the instrumental effects can be accurately represented in parametric
form, giving reliable estimates that cannot be obtained from a deconvolved image alone. We devise a model
with adjustable parameters, which we believe is capable of describing the data, and choose a \textit{figure-of-merit} or \textit{merit 
function} to measure how well the 
data and the model agree with each other \citep[Chapter~15]{press2007}. In the Bayesian approach, the merit function is just the posterior
probability distribution and the data, in the present case, are the observed visibilities.

Statistical analysis of visibilities complements the traditional imaging and deconvolution techniques and, if applied judiciously, 
can improve on them. Sparse \textit{uv}-coverage, the spreading of localised \textit{uv}-domain errors throughout the image (thereby making them correlated
between pixels) by the Fourier transform process, 
and the difficulty in estimating the measurement uncertainties (or the \textit{noise}) from an image deconvolved using non-linear 
deconvolution techniques such as 
CLEAN \citep{hogbom1974} or MEM 
\citep{ables1974}, are but some of the factors that render Fourier-transform imaging difficult and often inadequate 
\citep[Chapter~16]{Syn1999}.

Our primary assumptions are that (i) the visibility measurements are independent and (ii) the uncertainties in the measurements follow 
a Gaussian distribution. If the data are independent, then the likelihood $\mathcal{L}$ from equation (\ref{eq:bayesparest}) can be 
expressed as \citep[Chapter~3]{sivia2006}
\begin{equation}
\label{eq:independence}
\mathcal{P}(D|\Theta,H) = \prod_{k=1}^{N_{\mathrm{vis}}} \mathcal{P}(D_k|\Theta,H)\quad,
\end{equation}
where $D_k$ stands for the $k$th datum and $N_{\mathrm{vis}}$ denotes the number of visibilities. Under the assumption that the visibility noise is 
Gaussian, the probability of obtaining each individual datum is given by
\begin{equation}
\label{eq:gaussian}
\mathcal{P}(D_k|\Theta,H) = \frac{1}{\sigma_k\sqrt{2\pi}}\exp{\left(-\frac{(F_k-D_k)^2}{2\sigma_k^2}\right)}\quad,
\end{equation}
where $\{\sigma_k\}$ denotes the expected uncertainties and $F_k \equiv f(\Theta,k)$ denotes a functional form of the model for a specific set of 
parameter values $\Theta$ for the $k^{\mathrm{th}}$ datum.

Combining equations (\ref{eq:independence}) and (\ref{eq:gaussian}), we obtain
\begin{equation}
\label{eq:maxlike}
\mathcal{P}(D|\Theta,H) \propto \exp \left(-\frac{\chi^2}{2}\right)\quad,
\end{equation}
where the sum of the squares of the \textit{normalised residuals} is represented by
\begin{equation}
\label{eq:chisquared}
\chi^2 = \sum_{k=1}^{N}\left(\frac{F_k-D_k}{\sigma_k}\right)^2\quad.
\end{equation}
If we also assume a flat (constant) prior for the parameters to indicate that we are largely ignorant of their expected values, 
then, taking the natural logarithm of equation 
(\ref{eq:bayesparest}) and omitting the normalisation constant, we arrive at
\begin{equation}
\mathrm{ln}[\mathcal{P}(\Theta|D,H)] = \mathrm{C}(\sigma_k) - \frac{\chi^2}{2}\quad.
\end{equation}
where $\mathrm{C}(\sigma_k)$ depends only on $\sigma_k$. Thus, with some simplifying assumptions, Bayesian parameter estimation 
reduces to the more familiar methods of \textit{maximum likelihood} and \textit{least-squares estimation}.
The crucial difference is that Bayesian inference is independent of the Gaussianity of the underlying process.
The exact form our likelihood function takes is very close to this equation (section \ref{subsec:likelihoodfn}). 

\subsection{The Measurement Equation}
\label{subsec:rimeandsw}
The simultaneous estimation of source and instrumental parameters is facilitated by the \textit{Radio Interferometer Measurement
Equation} (RIME), originally formulated by \citet{hbs1996} and reformulated for \textit{direction-dependent effects} (DDEs) 
by \citet{oms2011}. 
The RIME unifies the concept of the Stokes parameter representation of electromagnetic waves \citep{BornWolf1999} and the technique of radio 
interferometry under a single mathematical framework.

For a sky composed of discrete point sources, the RIME may be written as
\begin{equation}
\label{eq:rime1}
\mathbfss{V}_{pq} = \sum_s \mathbfit{J}_{sp}\, \mathbfss{B}_s\, \mathbfit{J}_{sq}^H\quad,
\end{equation}
where $\mathbfss{V}_{pq}$ denotes the $2 \times 2$ matrix of visibilities measured by the baseline $pq$ with contribution from each 
discrete source $s$, 
$\mathbfit{J}_{sp}$ is a \textit{Jones matrix} that incorporates all propagation path effects on the way from the source $s$ to the antenna $p$, 
the superscript $H$ denotes the Hermitian conjugate, and
\editor{$\mathbfss{B}_s$ is the brightness matrix} of source $s$, which relates the correlated voltages from the two polarisation feeds to the four Stokes 
parameters $I, Q, U,$ and $V$\footnote{Here, we assume that the \referee{two feeds are sensitive to circular
polarisation; for linear polarisation}, the brightness
matrix undergoes a further linear transformation \citep[section 6.3]{oms2011}.}:
\begin{equation}
\label{eq:brightnessmatrix}
\mathbfss{B}_s = \begin{pmatrix} I+V && Q+iU \\ Q-iU && I-V \end{pmatrix}\quad.
\end{equation}
In the more general case,
if we consider the sky to be a continuous brightness distribution, the RIME may be written as
\begin{equation}
\label{eq:rime2}
\mathbfss{V}_{pq} = \int\displaylimits_{4\pi} \mathbfit{J}_p(\bm{\widehat{\sigma}})\, \mathbfss{B}(\bm{\widehat{\sigma}})\, \mathbfit{J}_q^H(\bm{\widehat{\sigma}})\quad \mathrm{d}\Omega\quad,
\end{equation}
where $\bm{\widehat{\sigma}}$ represents the unit direction vector and $d\Omega$ denotes spherical integration over the entire sky. The known 
DDEs and DIEs (\textit{direction-independent effects}) may each be assigned its own Jones matrix, while the unknown 
effects may be subsumed into a generic Jones matrix term. This approach is called the \textit{phenomenological} RIME and has been implemented in 
the MeqTrees\footnote{http://meqtrees.net} software suite for interferometric simulation and calibration \citep{meqtrees2010} that we use for
our simulations (section \ref{sec:martividal}).
To interface with M\textsc{ulti}N\textsc{est}, which samples the posteriors and computes the evidence, we use 
PyMultiNest\footnote{https://johannesbuchner.github.io/PyMultiNest}, a python wrapper to M\textsc{ulti}N\textsc{est} \citep{pymultinest2014}.

Unlike the toy model selection problem presented in \citet{biro2015}, the models we use here incorporate instrumental systematics and hence contain
more parameters (section \ref{subsec:models}).
To evaluate the likelihood and consequently, estimate the evidence, the RIME must be computed at each iteration of the Bayesian inference
process. To accurately model the visibilities given a sparse sky model, the RIME is implemented using the computationally expensive direct 
Fourier transform. 
We use \textit{Montblanc}\footnote{https://github.com/ska-sa/montblanc} \citep{mb2015}, 
a GPU-accelerated implementation of the RIME for this. Montblanc is built on PyCUDA, a Python interface to NVIDIA's CUDA architecture 
\citep{pycuda2012}, to parallelise the RIME evaluation. 
Montblanc computes the RIME from sampled model parameters during each iteration to generate the $\chi^2$ value 
(equation \ref{eq:chisquared}), which may be used for calculating the likelihood. 
Currently, it supports the simulation of three source morphologies (point, Gaussian, and S\'ersic) 
of which we use the point and Gaussian source models, and the inclusion of DIEs and DDEs in the RIME.

\section{Data Description and reduction}
\label{sec:dataanalysis}
Blazars are galaxies that exhibit strong radio emission from their cores (called \textit{active galactic nuclei} (AGN)) in the form 
of jets of relativistically beamed particles along the line of sight to the observer \citep[e.g.,][]{antonucci1993, urry1995}. 
They display high radio luminosities (10$^{46}\: $erg s$^{-1}$) generated by non-thermal
radiation mechanisms, mainly synchrotron and inverse Compton radiation \citep{rybicki1985}, with flat spectra. Owing to their compactness, 
blazars are an important class of objects whose studies have largely benefited from VLBI observations.

\begin{figure}
 \includegraphics[width=\columnwidth]{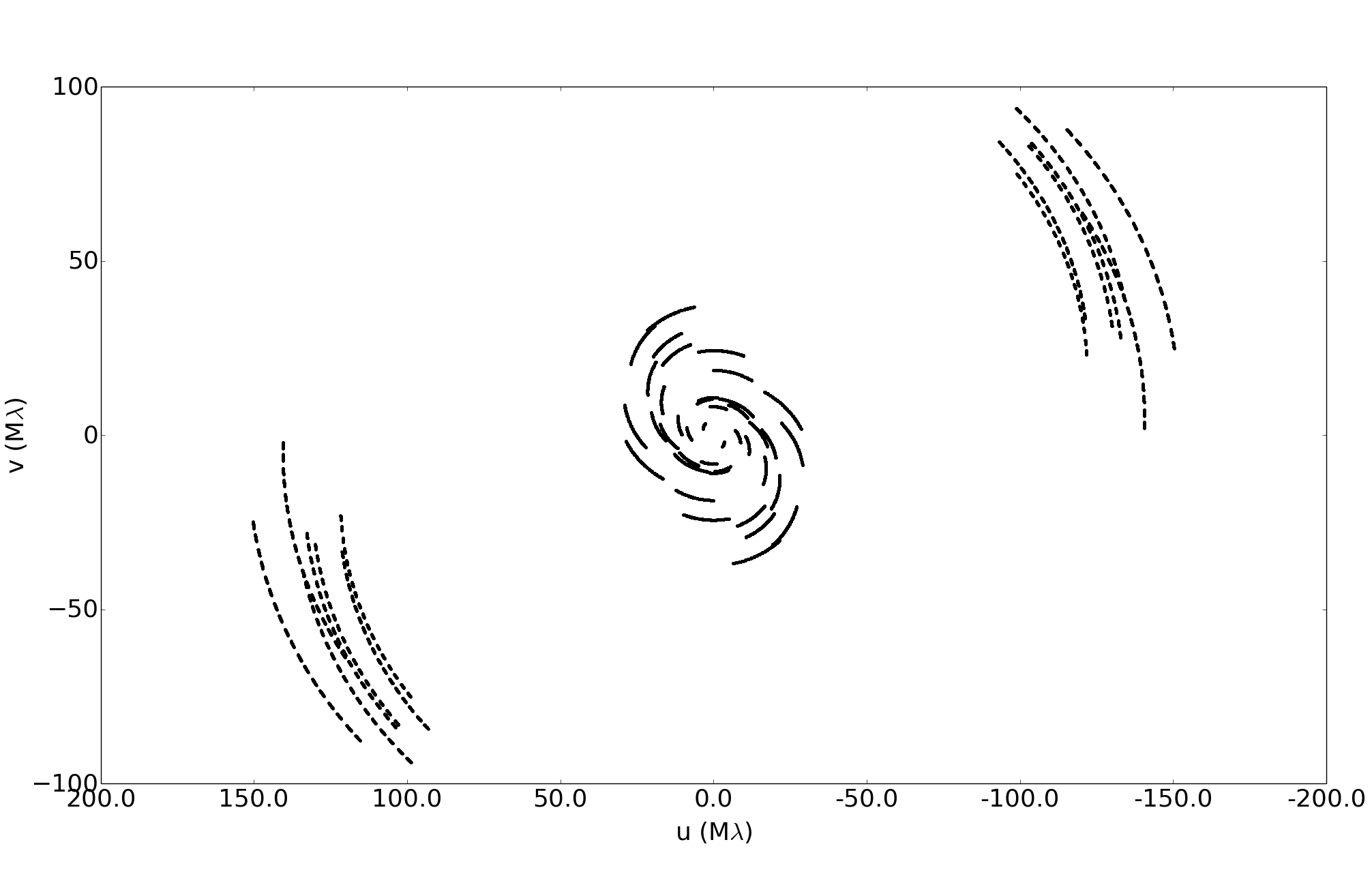}
 \caption{The \textit{uv}-coverage of the synthesis.}
 \label{fig:uvcov}
\end{figure}

J0809+5341 is located at J$\, 08^\mathrm{h}09^\mathrm{m}41.733^\mathrm{s},\, +53^\mathrm{d}41^\mathrm{m}25.092^\mathrm{s}$ 
at redshift $z=2.144$ \citep{sourceredshift2014}. It was observed using 8 EVN stations 
(Table \ref{tab:stations}), for $130$ minutes using $8 \times 16$ MHz spectral bands of $32$ channels each, between $4926$ and 
$5054$ MHz, with $2$ s integration time, in both R and L polarisations. 
\begin{table}
\caption{EVN stations used in the observation along with the corresponding dish diameters and the nominal (SEFD) values.}
\begin{center}
\begin{tabular}{cccc}
\hline
Station  & Code & Diameter (m) & Nominal SEFD (Jy)\\
\hline
Effelsberg & EF & 100 & 20\\
Jodrell Bank & JB & 25 & 320\\
Noto & NT & 32 & 260\\
Onsala & ON & 25 & 600\\
Torun & TR & 32 & 220\\
OAN-Yebes & YS & 40 & 160\\
Westerbork & WB & 25 & 120\\
Sheshan & SH & 25 & 720\\
\hline
\end{tabular}
\end{center}
\label{tab:stations}
\end{table}
The \textit{uv}-coverage is shown in Figure \ref{fig:uvcov}. The long baselines correspond to the Sheshan station in Shanghai (SH) and extend
over 9000 km. Without SH, the maximum baseline length is about 2200 km.
More details on this project can be found in \citet{antao2016}.

\referee{Invalid data were excluded (or \textit{flagged}) before the visibilities were \textit{self-calibrated}
\citep{selfcal1978}\footnote{\updated{www.evlbi.org/user\_guide/evn\_datareduc.html}}.
Self-calibration is the process of introducing some plausible assumptions about the source structure to correct the 
observed complex visibilities \citep{Syn1999}.}
For this procedure, the channels in each band were averaged together so that the data 
contained 8 bands of one channel each. 
\updated{This frequency-averaging is made possible by the fact that the source is located
at the pointing centre and is therefore not affected by frequency smearing \citep{oms2011}.}
We extracted the spectral band centred at 4982.24 MHz for our analysis. 
The original UVFITS file was converted into the Measurement Set (MS) format and the missing\footnote{Due to the source not being visible 
to those baselines.} baselines were introduced -- with the corresponding data flagged -- for compatibility with Montblanc.

\begin{figure}
\center
 \includegraphics[scale=0.24]{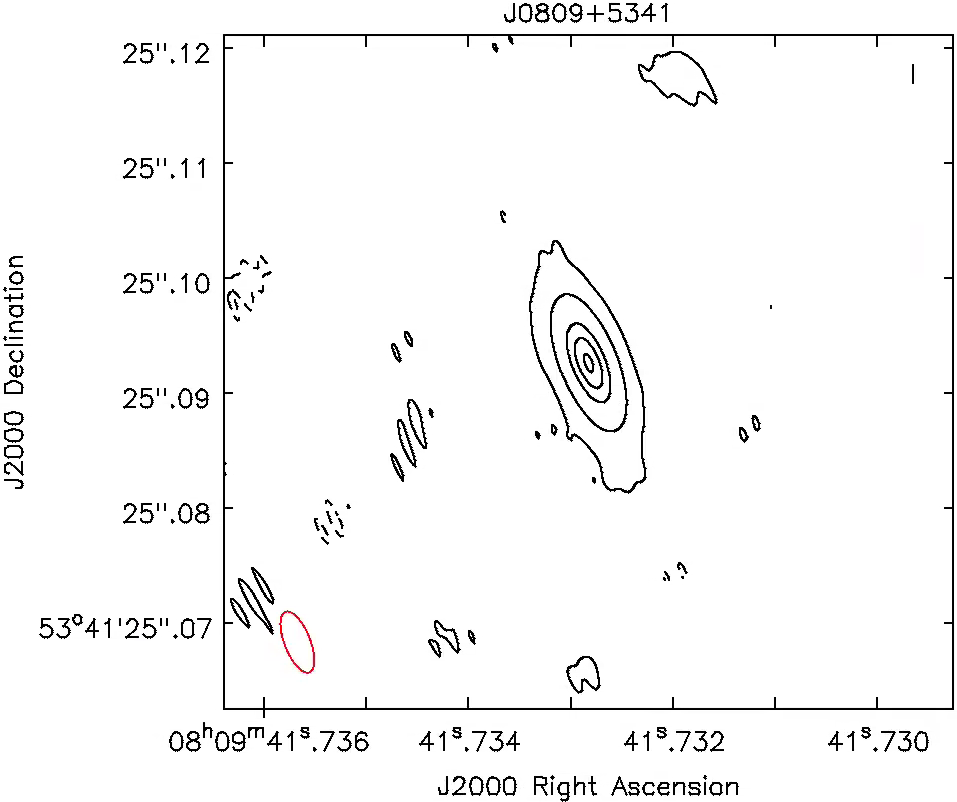}
 \caption{Stokes $I$ contour image of J0809+5341. 
The contours presented are $-3, 3, 100, 1000, 2000$, and $3000$ times the rms noise in the image ($\sim0.05$ mJy), with the negative contours shown as dashed lines. 
The red ellipse at the bottom left corner is the PSF used for restoration by CLEAN: $5.7 \times 2.2$ mas, oriented at an angle of $21.7^\circ$.}
 \label{fig:image-naturalwt}
\end{figure}

A \textit{naturally}-weighted \citep{danbriggs}, 
deconvolved image of the self-calibrated data shows a compact source (Figure \ref{fig:image-naturalwt}). 
\referee{The deconvolution was performed using the \textsc{csclean} algorithm \citep{csclean1984} in
\textit{lwimager}\footnote{\referee{https://github.com/casacore/casarest}} with 1000 iterations. The point spread function (PSF) of the 
interferometer 
was calculated using the Hogbom algorithm recommended for data with poor \textit{uv}-coverage \citep{clean1974}.}
The dynamic range of the image is 3200:1. Image-plane source extraction using PyBDSM\footnote{http://www.astron.nl/citt/pybdsm} 
estimates that the source is unresolved with a flux density of $161.1\pm0.01$ mJy.
\referee{In comparison, \citet{antao2016} performed model-fitting on the data using \textit{Difmap}\footnote{\referee{ftp://ftp.astro.caltech.edu/pub/difmap/difmap.html}}, 
a software package for self-calibrating visibilities and iteratively building up a model of the sky \citep{shepherd1997}, and 
concluded that the best-fit model was a circular Gaussian of size $0.31\pm0.06$ mas.}

\section{Bayesian Analysis}
\label{sec:bayesblazar}
\subsection{Description of Models}
\label{subsec:models}
We compare three morphological models to describe unresolved or partially-resolved sources, each differing in the assumed
brightness distribution of the source:
\begin{description}
\item [\textbf{PT:}] The point source model consists of one flux density parameter, $S_{\nu}$, and two parameters
that describe the position of the source, $(l, m)$, the direction cosines measured with respect to the ($u, v$) co-ordinates.
\item [\textbf{GAU:}] The elliptical Gaussian model consists of $S_{\nu}$ and $(l, m)$ along with three more parameters, $l_p, m_p$, and $r$, 
describing the shape of the source. These three parameters are related to the major axis $e_{\mathrm{maj}}$, minor axis $e_{\mathrm{min}}$, and 
position angle or orientation $\theta$ of the ellipse as follows:
\begin{equation}
\begin{aligned}
l_p &= e_{\mathrm{maj}}\,\sin\theta \\
m_p &= e_{\mathrm{maj}}\, \cos\theta \\
r &= e_{\mathrm{min}} / e_{\mathrm{maj}}\quad. \\
\end{aligned}
\label{eq:shapepars}
\end{equation}
Thus $l_p$ and $m_p$ denote the projections of $e_{\mathrm{maj}}$, onto the $l$ and $m$ axes, and $r$ is the ratio of the minor axis
to the major axis (see Figure \ref{fig:gaussshape}).
\begin{figure}
\center
 \includegraphics[scale=0.2]{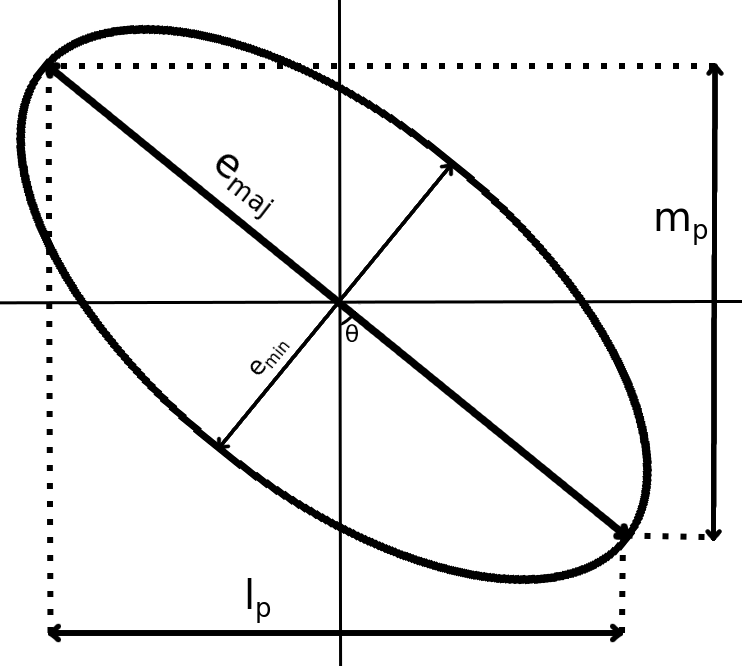}
 \caption{The relation between the shape parameters, $l_p, m_p$, and $r$, and the major axis ($e_{\mathrm{maj}}$), 
minor axis ($e_{\mathrm{min}}$), and position angle ($\theta$) of an elliptical Gaussian profile.}
 \label{fig:gaussshape}
\end{figure}
\item [\textbf{CIRC:}] The circular Gaussian model consists of the same six parameters as GAU, with the exception that the ratio $r$ is given a
delta distribution at unity, forcing the shape to be circular.
\end{description}

A summary of the parametrisation of each model is given in Table \ref{tab:models}.
\begin{table}
 \caption{Models evaluated in this study. Besides the source parameters, there are 7 free parameters describing the station gain amplitudes
and 8 parameters describing the individual station SEFDs.}
 \label{tab:models}
 \begin{tabular}{ccc}
  \hline
  Model & Number of parameters & Parametrisation\\
  \hline
  \multirow{4}{*}{\textbf{PT}} & \multirow{4}{*}{18} & Flux Density ($S_{\nu}$) \\ 
  & & Position ($l, m$) \\ 
  & & Station gain amplitudes ($|g_p|$) \\ 
  & & $\mathrm{SEFD}_p$ \\ \hline
  \multirow{5}{*}{\textbf{GAU}} & \multirow{5}{*}{21} & Flux Density ($S_{\nu}$) \\ 
  & & Position ($l, m$) \\ 
  & & Shape ($l_p, m_p, r$) \\ 
  & & Station gain amplitudes ($|g_p|$) \\ 
  & & $\mathrm{SEFD}_p$ \\ \hline
  \multirow{5}{*}{\textbf{CIRC}} & \multirow{5}{*}{20} & Flux Density ($S_{\nu}$) \\ 
  & & Position ($l, m$) \\ 
  & & Shape ($l_p, m_p$) \\ 
  & & Station gain amplitudes ($|g_p|$) \\ 
  & & $\mathrm{SEFD}_p$ \\
  \hline
 \end{tabular}
\end{table}
Alongside the parameters that describe the source, all three models incorporate the same instrumental parameters. 
\referee{In principle, any instrumental effect could be modelled using the RIME (see section \ref{subsec:rimeandsw}), while in practice, we are often
limited by the software used to sample the parameter space. For instance, M\textsc{ulti}N\textsc{est} performs best with low-dimensional 
($\lesssim30$) parameter spaces. For the present experiment, the most important parameters that affect the source shape resolution are the amplitudes
of the complex station gains and the \textit{System Equivalent Flux Densities} (SEFD) of each antenna. 
The complex station gains vary with respect to time during the course of the observation and are solved for during the preliminary 
self-calibration. The resulting gain amplitudes and phases 
are sufficiently smooth to be modelled using one complex gain term per station. 
Moreover, since all the source models considered place the source at the pointing centre, the measured phases on all
the baselines will be zero. Hence, we hold the gain phases constant at zero.}
The SEFD is the flux density equivalent of the system temperature $T_{sys}$ of a receiver system \citep{TMS2001}. It varies 
widely between stations in any VLBI observation and is here used to derive the per-visibility noise term $\sigma_{pq}$ for each baseline 
$pq$, which are then 
used to weight the model visibilities using the inverse of the corresponding variance. 
This brings down the number of parameters that describe the noise from the \textit{number of baselines} to the \textit{number of stations}.
The per-visibility uncertainty for one polarisation in terms of the geometric mean of the station SEFDs is given by the 
\referee{\textit{radiometer equation} \citep{TMS2001}:}
\begin{equation}
\label{eq:varyingsigma}
\begin{aligned}
\sigma_{pq} &= \frac{\mathrm{SEFD}_{pq}}{\sqrt{\delta\nu\, \tau_{pq}}}\quad, \\
\mathrm{where}\ \mathrm{SEFD}_{pq} &= \sqrt{\mathrm{SEFD}_p\, \mathrm{x}\, \mathrm{SEFD}_q}\quad, \\
\end{aligned}
\end{equation}
$\mathrm{SEFD}_p$ is the SEFD of station $p$, $\delta\nu$ is the channel bandwidth, and $\tau_{pq}$ is the integration time for
baseline $pq$.

\subsection{Building the RIME}
\label{subsec:ourrime}
We are now ready to construct the RIME for modelling the visibilities. 
Assuming a flat spectral index and using the flux density parameter $S_{\nu}$ in the brightness matrix (equation \ref{eq:brightnessmatrix}),
we obtain
\begin{equation}
\label{eq:ourrimeb}
\mathbfss{B} = \begin{pmatrix} S_{\nu} && 0 \\ 0 && S_{\nu} \end{pmatrix}\quad.
\end{equation}
The first linear transformation this signal undergoes is represented by the phase delay matrix $K$, associated with the difference 
in the geometric path lengths from the source to antennas $p$ and $q$. Given the phase difference ($\kappa_p$) between the waves received by antenna $p$ 
located at $\mathbf{u}_p = (u_p, v_p, w_p)$ relative to $\mathbf{u}=0$, the scalar \textit{K-Jones} term for antenna $p$ can be written as
\begin{equation}
\label{eq:kjones}
K_p = \mathrm{e}^{-i\kappa_p} \equiv \mathrm{e}^{-i\kappa_p} \begin{pmatrix} 1 && 0 \\ 0 && 1 \end{pmatrix}\quad.
\end{equation}
The K-Jones term must be accounted for even under ideal conditions in which nothing else affects the signal from the source to the 
interferometer. Knowing this, we may define the \textit{source coherency} matrix as \citep{oms2011}
\begin{equation}
\label{eq:soucoh}
\mathbfss{X}_{pq} = K_p\, \mathbfss{B}\, K_q^H = \mathbfss{B}\mathrm{e}^{-i\kappa_{pq}}\quad.
\end{equation}
If we assume equal gains for the two polarisation feeds, then $g_p = g_{x_p} = g_{y_p}$, and the diagonal \textit{G-Jones} 
matrix describing station gains reduces to a scalar matrix,
\begin{equation}
\label{eq:gjones}
G_p = \begin{pmatrix} g_p && 0 \\ 0 && g_p \end{pmatrix} = g_p\, \begin{pmatrix} 1 && 0 \\ 0 && 1 \end{pmatrix}\quad.
\end{equation}
The PSF of the observation is $5.7 \times 2.2$ mas when the visibilities are weighted naturally, while the primary beams of the 
stations are at least 8 arcmin wide at 5 GHz. 
Hence, for a source located at the pointing centre, the \textit{E-Jones} matrix for primary beam effects may be set to unity, 
further simplifying the RIME.

Now, \editor{remembering the additive Gaussian noise term with zero mean and a variance of $\sigma_{pq}^2$ per visibility}, the RIME for the point source becomes
\begin{equation}
\label{eq:ourrimept}
\mathbfss{V}_{pq} = G_p\, \mathbfss{X}_{pq}\, G_q^H + \mathcal{N}(0, \sigma_{pq}^2)\quad.
\end{equation}
For the extended source models, the brightness distribution is integrated over the extent of the source. Expressed in terms of the direction cosines $l$ and $m$, 
the RIME for the extended source may be written as \citep[section 3.1]{TMS2001}
\begin{equation}
\begin{aligned}
\label{eq:ourrimegauss}
\mathbfss{V}_{pq} &= \iint\displaylimits_{lm}G_p\, \mathbfss{X}_{pq}(l, m)\, G_q^H\: \mathrm{d}\Omega\ +\ \mathcal{N}(0, \sigma_{pq}^2), \\
\mathrm{where}\,\,\mathrm{d}\Omega &= \frac{\mathrm{d}l\, \mathrm{d}m}{\sqrt{1-l^2-m^2}}\quad.
\end{aligned}
\end{equation}

\subsection{Likelihood function}
\label{subsec:likelihoodfn}
Now that we have described the models quantitatively, we can set up the likelihood function for the problem at hand. 
Following the discussion in section \ref{subsec:statsanalysis}, given the observed ($V_{D}$) and
the modelled ($V_M$) visibilities, and the uncertainties $\sigma_{k}$ that vary with baseline, the likelihood function for parameter
estimation for model $H$ may be written as
\begin{equation}
\begin{aligned}
\label{eq:ourlikelihood}
\mathcal{L}(\Theta|V_D,H) &= \frac{1}{\displaystyle\prod_{k=1}^{2N_{\mathrm{vis}}}\, \sqrt{2\pi \sigma_k^2}} \exp \left( -\frac{\chi^2}{2} \right)\quad, \\
\mathrm{where}\ \chi^2 &= \sum_{k=1}^{2N_{\mathrm{vis}}}\left(\frac{V_{M_k}-V_{D_k}}{\sigma_k}\right)^2\quad,
\end{aligned}
\end{equation}
and $N_{\mathrm{vis}}$ is the total number of complex visibilities. The summation is carried out over $2N_{\mathrm{vis}}$ 
since we consider the real and imaginary parts separately. Taking the natural logarithm of $\mathcal{L}$, we obtain
\begin{equation}
\begin{aligned}
\label{eq:logl}
\mathrm{ln}(\mathcal{L}) &= \sum_{k=1}^{2N_{\mathrm{vis}}} \mathrm{ln} \left[ (2\pi\sigma_k^2)^{-1/2} \right] - \frac{\chi^2}{2} \\
&= -\frac{1}{2} \sum_{k=1}^{2N_{\mathrm{vis}}} \mathrm{ln} \left[ 2\pi\sigma_k^2 \right] - \frac{\chi^2}{2}\quad. \\ 
\end{aligned}
\end{equation}
The noise is modelled such that the variance for the real and imaginary parts of a complex visibility is the same. 
Hence, counting each $\sigma_k$ twice, we arrive at the final form of the log-likelihood function:
\begin{equation}
\label{eq:ourloglikely}
\mathrm{ln}(\mathcal{L}) = - \sum_{k=1}^{N_{\mathrm{vis}}} \mathrm{ln} \left[ 2\pi\sigma_k^2 \right] - \frac{\chi^2}{2}\quad.
\end{equation}

\subsection{Prior distributions}
\begin{table}
 \center
 \caption{Prior distributions for the different parameters used. All the listed parameters were set uniform priors with the range indicated by the values in
the square brackets. The parameters with delta priors are not included.}
 \label{tab:priors}
 \begin{tabular}{ll}
  \hline
  Parameter (units) & Prior distribution \\
  \hline
  $S_{\nu}$ / Jy & [0.1, 0.2] \\
  $l \& m$ / mas & [-4, 4]\\
  $l_p$ / mas & [0, +4] \\
  $m_p$ / mas & [-4, +4] \\
  $r$ & [0, 1] \\
  $|g_p|$, where $p\ne\mathrm{EF}$ & [0.8, 1.2] \\
  SEFD / Jy & [5, 800] \\
  \hline
 \end{tabular}
\end{table}
We set uniform (flat) priors $\Pi(\Theta)$ on most parameters (Table \ref{tab:priors}). The prior range for $S_{\nu}$ is chosen based on what we know about 
the flux density
of J0809+5341 from preliminary imaging and source extraction (section \ref{sec:dataanalysis}). The prior distribution for $l_p$ 
(the sine projection of $e_{\mathrm{maj}}$) is restricted to non-negative values so that the position angle estimate is constrained to a range of $180^{\circ}$. 
We allow the gain amplitudes to vary between $\pm20$ per cent of unity and keep the corresponding phases fixed at 
zero\footnote{\updated{This is achieved by setting delta distributions centred at the known values of the parameters. The data do not impact the posteriors of such 
parameters since their posteriors are also delta distributions at the same locations in the parameter space.}}.
To break the degeneracy between $S_{\nu}$ and $|g_p|$ estimates, the gain amplitude of the EF station is set a delta prior at unity; we choose EF because it provides some 
of the shortest baselines (except for the baseline with SH) in the synthesis and is not sensitive to the structure of the 
compact source, and the preliminary self-calibration tells us that EF has the most stable gain.

For the model selection step, we assign equal priors to all the models considered, so that the logarithm of the Bayes factor $B_{12}$ 
may directly be used for model comparison.

\section{Results}
\label{sec:results}
The M\textsc{ulti}N\textsc{est} sampler takes the prior distributions and the likelihood as its inputs, computes the natural logarithm
of the Bayesian evidence (equation (\ref{eq:evidence})) for each model, and produces 
the joint posterior as a by-product. We evaluate each model independently and use the corresponding logarithmic evidences
to compute the Bayes factor between models using equation (\ref{eq:logbayesfactor}). 

The analysis reveals that the factors $2\, \mathrm{ln} (B_{ij})$, for GAU against PT and CIRC are 
$21.0\pm0.8$ and $26.0\pm0.8$ respectively. Applying the criteria we set aside in
Table \ref{tab:evidences}, we conclude that there is a \textit{very strong} preference for the elliptical Gaussian (GAU) over the 
other two models\footnote{The quoted relative ln-evidences are obtained from the \textit{importance nested sampling} results owing to
their better accuracy, while the uncertainties are obtained from the \textit{vanilla} nested sampling results \citep{multinest3}.}.

\begin{figure}
\center
 \includegraphics[scale=0.26]{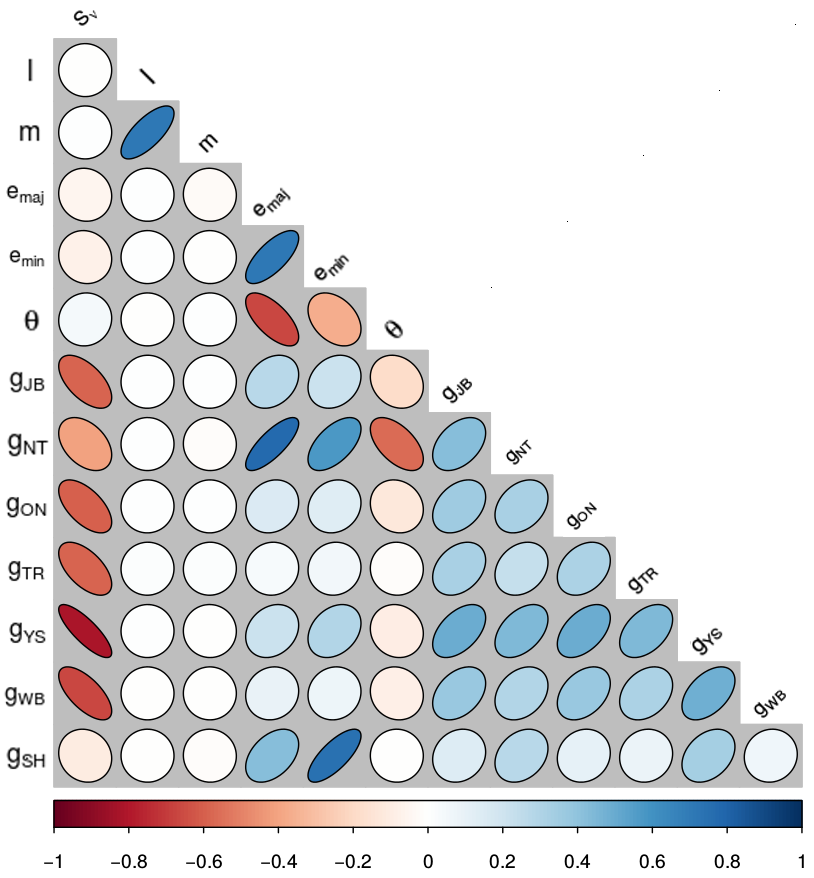}
 \caption{Correlations between the estimated parameters for model GAU.
The quantities $e_{\mathrm{maj}}$, $e_{\mathrm{min}}$, and $\theta$ are derived from the shape parameters. 
\referee{The individual station SEFDs and parameters with delta prior distributions are excluded from the figure}. The coloured ellipses correspond to the correlation coefficients
shown at the bottom; the higher the ellipticity, the stronger the correlation.}
 \label{fig:corrs}
\end{figure}

Figure \ref{fig:corrs} shows the correlations between various parameters of GAU.
We use the quantities $e_{\mathrm{maj}}$, $e_{\mathrm{min}}$, and $\theta$, obtained from equations (\ref{eq:shapepars}), for visualisation.
\referee{The individual station SEFDs are mostly uncorrelated with the other parameters and with each other, and are not shown in the figure}.
The station gain amplitudes are correlated negatively with the estimated flux density as one would expect: the higher the instrument
gain, the lower the true flux density of the source.
The gain amplitude, $|g_{\textsc{sh}}|$, of SH, which provides the longest baselines, is correlated positively with the shape parameter estimates.

\begin{figure*}
\center
 \includegraphics[scale=1.05]{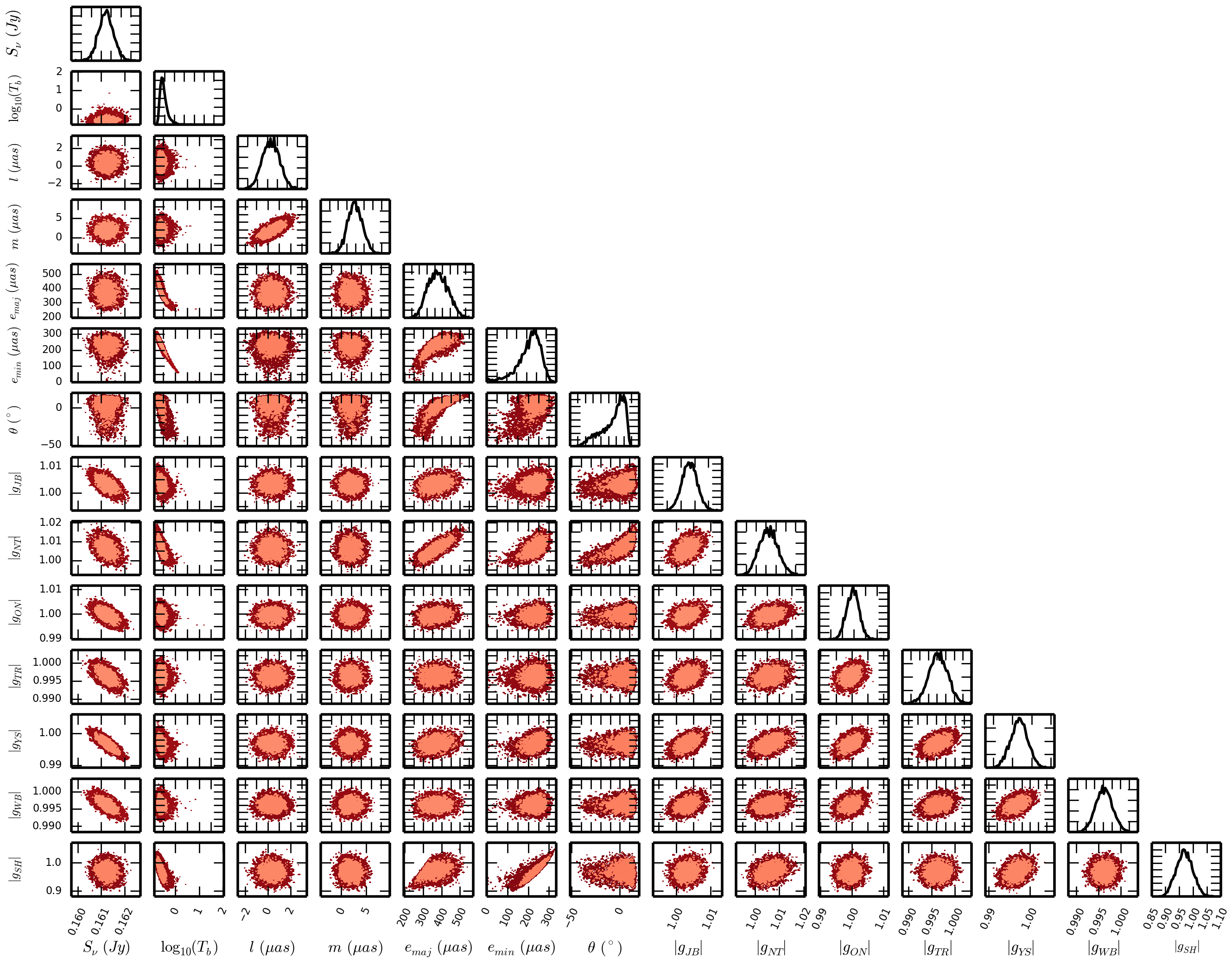}
 \caption{The 1-D posteriors and 2-D correlations of the source and the station gain parameters for model GAU. The principal diagonal gives the 1-D marginalised 
posterior distributions of the estimated flux density ($S_{\nu}$), 
the source brightness temperature ($T_b$), the derived shape quantities ($e_{\mathrm{maj}}$, $e_{\mathrm{min}}$, $\theta$), 
and the station gain amplitudes $|g_p|$, while the lower triangular matrix gives the 2-D joint posteriors
between their various combinations. 
The 68 and 95 per cent credible regions are indicated by the light-red and dark-red shaded regions respectively. 
Parameters with delta priors and the station SEFDs (which are uncorrelated with the other parameters) are excluded from the figure.}
 \label{fig:triangle-plot}
\end{figure*}

\begin{figure}
\center
 \hspace*{-0.3cm}\includegraphics[width=\columnwidth]{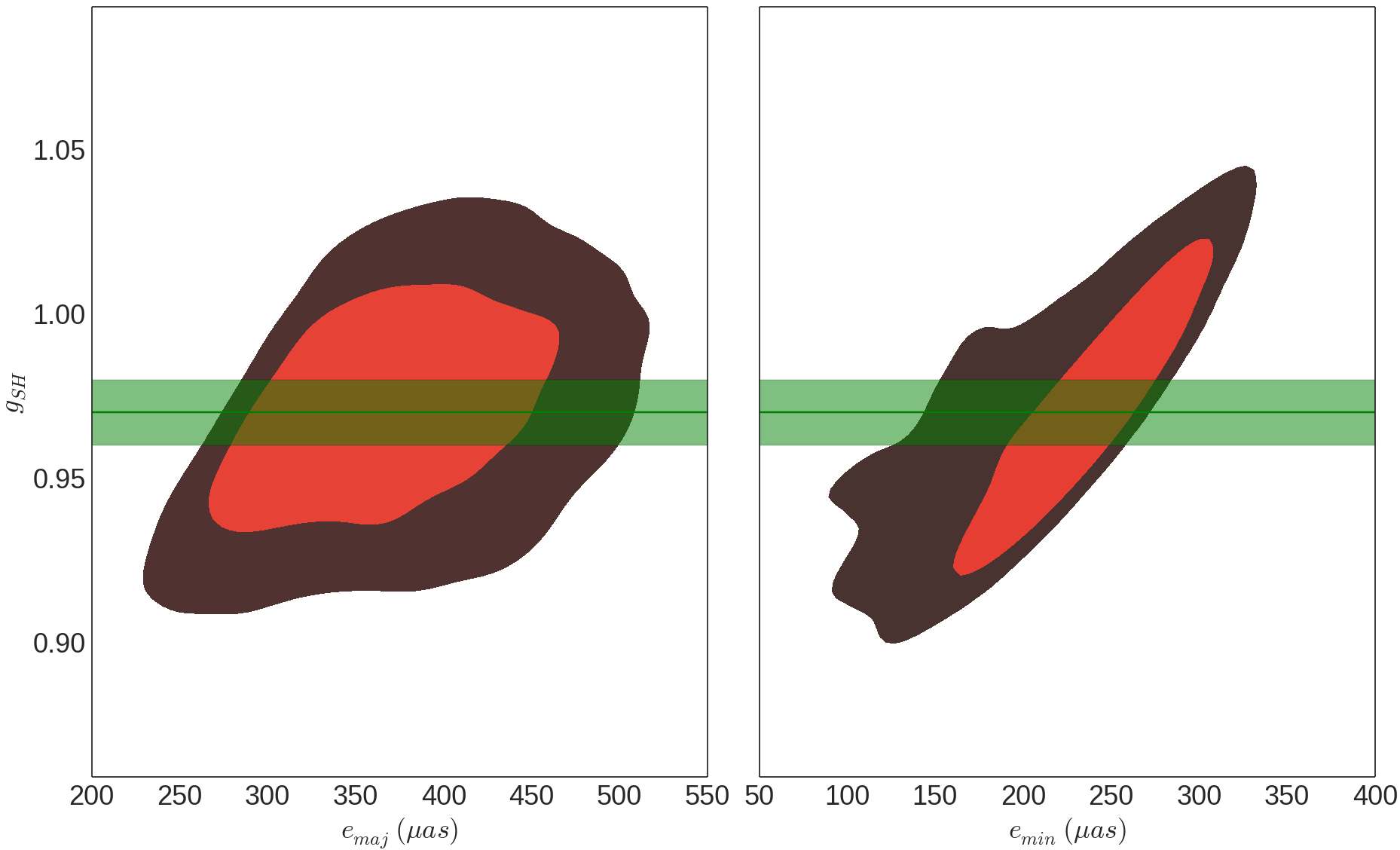}
 \caption{The correlations between SH gain amplitude and the shape parameters. The horizontal bars on either side of the regions shaded green 
correspond to $\pm$$1$ per cent about the mean of the marginalised posterior of $g_{\textsc{sh}}$. The light and dark red shaded regions indicate
the 68 and 95 per cent credible regions respectively of the posteriors of the shape parameters $e_{\mathrm{maj}}$ and $e_{\mathrm{min}}$.
It is evident that a better gain calibration would constrain both $e_{\rm min}$ and $e_{\rm maj}$, due to the strong correlation 
between $e_{\rm maj}$ and $e_{\rm min}$ (not highlighted here), better.}
 \label{fig:zoomedcorrs}
\end{figure}

Figure \ref{fig:triangle-plot} shows the 1-D and 2-D marginalised posteriors of the source and the station gain parameters, along with the
source brightness temperature, $T_b$ (section \ref{subsec:science}). 
The 2-D marginalised posterior distributions between $|g_{\textsc{sh}}|$ and the three shape parameters show
the precise nature of the relationships between them. Presenting these relationships in full is the most complete statement we can make about these parameters. 
The SH gain amplitude is not constrained to the same precision as those of the other stations by the preliminary self-calibration process.
\referee{SH does not form short baselines with any other station, thereby making it difficult for amplitude self-calibration to correct for the biased gains
\citep{martividal2012}.} Hence, measurements involving SH has 
contributions both from the resolved source structure and the uncertainties in the calibration of $|g_{\textsc{sh}}|$. 
Figure \ref{fig:zoomedcorrs} highlights the regions in the marginalised posteriors of 
$\, e_{\mathrm{maj}}$ and $e_{\mathrm{min}}$ that correspond to a 1 per cent variation in that of $|g_{\textsc{sh}}|$ (which varies by about 12 per cent) 
about its mean.
\referee{Providing short baselines for outlying stations such as SH by introducing more VLBI antennas close by would reduce the
sparsity of the array distribution and minimise the uncertainty in the 
estimation of the station gains. In Shanghai, there are now two radio telescopes available for VLBI: the old Sheshan 25m (SH) and the new Tianma 65m telescope 
which provides the necessary short baselines for SH \citep{kawaguchi2015}. When both are used in VLBI observations, the gain calibration of the 
longest EVN baselines will improve and result in reduced uncertainty in the derived shape parameter and brightness temperature distributions.}

There is also a strong correlation
between the position estimates (although with uncertainties of the order of $\mu$as), which may be due to the position not being 
constrained tightly enough along the extent of the major axis of the PSF. 

\subsection{European stations only}
The NT gain ampltude is positively correlated with the shape parameters (Figures \ref{fig:corrs} \& \ref{fig:triangle-plot}). In the absence of SH,
NT provides the longest baselines for the observation. This correlation motivated us to test whether the source can successfully be resolved using 
only the baselines pertaining to the European stations. The relative evidence between PT and GAU comes out to $2.86\pm0.75$, while that between PT 
and CIRC is $1.61\pm0.76$. There is positive evidence for PT and a mild preference for CIRC over GAU and so, without the SH measurements, we
are forced to conclude that the source is barely resolved. 

\subsection{Comparison with \textsc{difmap} results}
\label{subsec:difmap}
\referee{We also performed model-fitting to the data using Difmap, conventionally used in VLBI, so that we could compare our results with those 
reported by \citet{antao2016}.}
Difmap returns the best-fit parameter estimates and the reduced chi-squared, $\chi^2_{\mathrm{red}}$, for each model 
as a measure of its goodness-of-fit\footnote{$\chi^2_{\mathrm{red}} = \chi^2\, /\, \mathrm{DoF}$, where DoF stands for
\textit{degrees of freedom}, obtained by subtracting the number of model parameters from the number of measurements.}.
The $\chi^2_{\mathrm{red}}$ values for these models, with and without the SH baselines, are shown in Table \ref{tab:difmap}.
\begin{table}
\caption{Reduced chi-squared values for the three models with and without the SH measurements using Difmap. 
The asterisk indicates that these models do not take instrumental effects into account.}
\begin{center}
\begin{tabular}{ccc}
\hline
\multirow{2}{*}{Model} & $\chi^2_{\mathrm{red}}$ with SH & $\chi^2_{\mathrm{red}}$ without SH \\
& (DoF = 164173) & (DoF = 121501) \\
\hline
PT* & 2.2442 & 1.7266 \\
GAU* & 1.8746 & 1.7253 \\
CIRC* & 1.8785 & 1.7256 \\ 
\hline
\end{tabular}
\end{center}
\label{tab:difmap}
\end{table}
These models include only the source parameters and do not account for the
instrumental effects considered in the original three models (Table \ref{tab:models}). As a result, they are much simpler than the models we 
test using the RIME. 

With the SH measurements included, there is a slight preference for the resolved source models.
Without the SH measurements, as with the Bayesian analysis, PT is at least as good a fit as any other model if not better, because of its simplicity.
Though $\chi^2_{\mathrm{red}}$ has traditionally been used for model selection, it is a poor substitute for Bayesian evidence since it 
assumes that the underlying processes are Gaussian. The Bayesian approach does not depend on this assumption. Moreover, there is no 
measurement of the uncertainty in the value of $\chi^2_{\mathrm{red}}$, which becomes important when the $\chi^2_{\mathrm{red}}$ values of 
two models do not differ much. In contrast, the Bayesian evidence has an uncertainty associated with it which we may use to determine 
the significance of the model selection ratio.

\begin{figure}
\center
 \hspace*{-0.3cm}\includegraphics[width=\columnwidth]{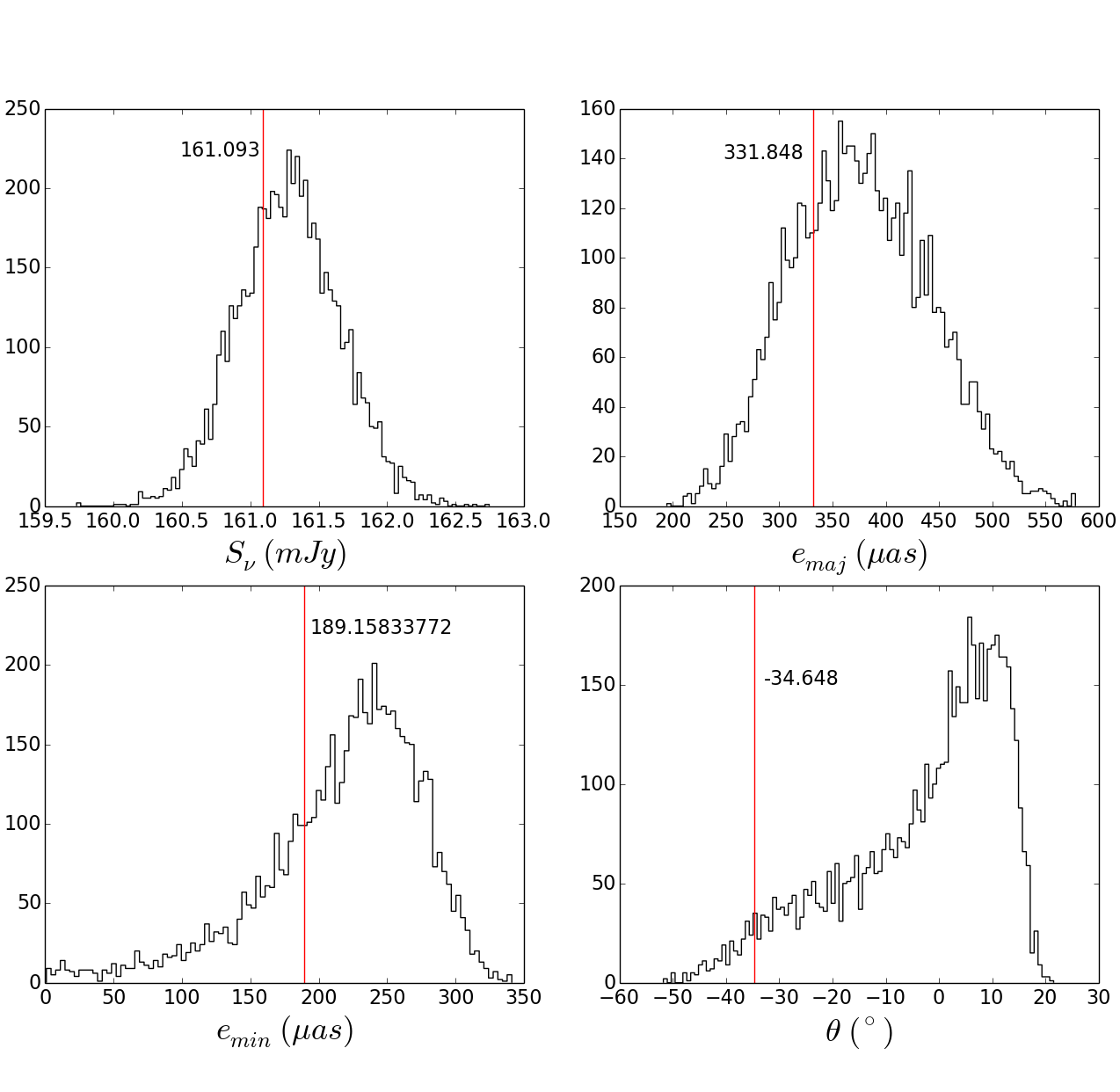}
 \caption{Comparison of the posteriors of the source parameters with the corresponding Difmap estimates. The vertical red lines correspond to the 
best-fit Difmap values printed alongside and the histograms correspond to the 1-D marginalised posteriors of the source parameters.}
 \label{fig:difmap}
\end{figure}

Figure \ref{fig:difmap} shows the relation between the Difmap best-estimates and the posterior distributions of the source
parameters obtained using the Bayesian analysis.
The Difmap estimates are point estimates with no associated uncertainties. They fall within the 68 per cent credible region of the Bayesian posteriors
when the posteriors are Gaussian ($S_{\nu}$ and $e_{\mathrm{maj}}$). For $e_{\mathrm{min}}$ and $\theta$, the posteriors are non-Gaussian and must be 
presented and accounted for in full while drawing inferences, such as computing the brightness temperature distribution of the blazar.

\subsection{What does this mean for the blazar?}
\label{subsec:science}

\begin{figure}
\center
 \includegraphics[width=\columnwidth]{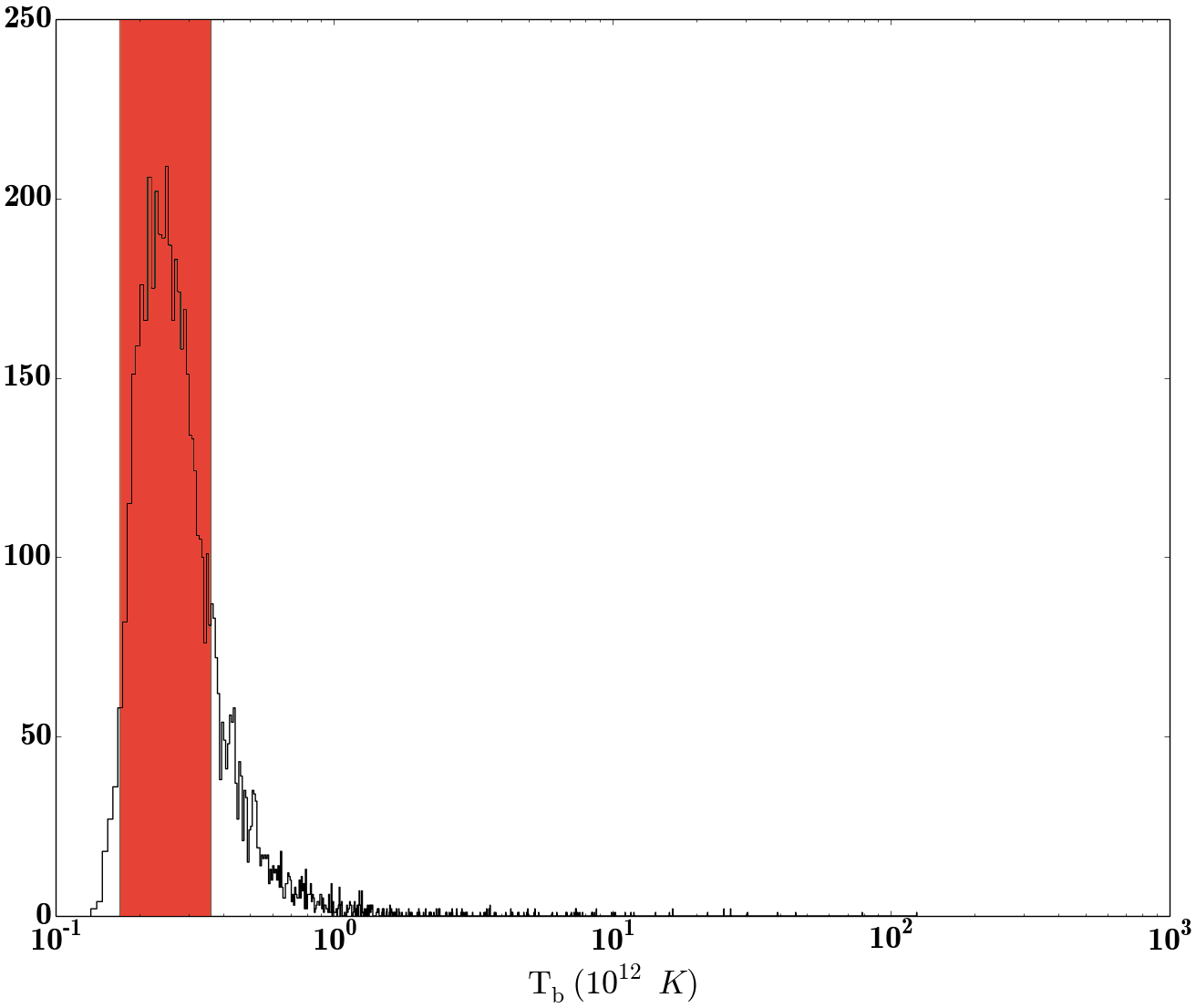}
 \caption{Histogram of the brightness temperature of J0809+5341 shown in black. The 68 per cent credible region (0.17 to 0.36) around the mode
(0.25) is shaded light red.}
 \label{fig:Tb-hist}
\end{figure}

In the radio regime, the brightness temperature $T_b$ of a source is given by the Rayleigh-Jeans approximation to Planck's law 
as \citep{Tools2009}
\begin{equation}
\label{eq:rj}
T_b = S_{\nu}\frac{c^2}{2k\nu^2}\, \frac{1}{\Delta\Omega}\quad,
\end{equation}
where $\nu$ is the frequency of observation, $k$ is the Boltzmann constant, and $\Delta\Omega$ is the solid area subtended by the source. 
For a Gaussian
source at high redshifts, such as J0809+5341, this relation becomes \citep{kellerman1988}
\begin{equation}
\begin{aligned}
\label{eq:tb}
T_b &= 1.22\, \frac{S_{\nu}}{\nu^2\, e_{\mathrm{maj}}e_{\mathrm{min}}}\, (1+z)\; 10^{12}\: \mathrm{K}\quad, \\
\end{aligned}
\end{equation}
With $\nu = 4.98224\, $GHz and $z = 2.144$, we use the full posterior distributions of $S_{\nu}$, $e_{\mathrm{maj}}$, and $e_{\mathrm{min}}$ 
to infer the distribution of $T_b$ (Figure \ref{fig:Tb-hist}). This ensures that our measurements account for the uncertainties in 
the source shape.
The mode of the distribution occurs at $0.25 \times 10^{12}\, \mathrm{K}$. The 68 per cent credible region places $T_b$ between
$0.17 \times 10^{12}\, \mathrm{K}$ and $0.36 \times 10^{12}\, \mathrm{K}$.
The few very high values of $T_b$ correspond to the very low values of $e_{\mathrm{min}}$.
 
\citet{rees1966} originally proposed that bright compact objects at cosmological distances are composed of parts moving at relativistic speeds.
This bulk relativistic motion is the most probable model that explains the high brightness temperatures and apparent superluminal motion of 
jets observed in radio galaxies \citep[e.g.,][]{marscher1980}.
An inverse Compton limit of $\simeq$$10^{11}$--$\, 10^{12}$ K for the brightness temperature was derived by \citet{kellermantoth1969}, which is the range 
in which our $T_b$ measurements for the blazar lie. This enhanced brightness temperature is a consequence of \textit{Doppler boosting} in 
which the relativistic beaming of a jet moving at an acute angle to the line of sight to the observer increases the observed flux density 
without increasing the size of the jet \citep{kellerman1988}.
The measured brightness temperature and the internal brightness temperature, $T_{b,int}$, relate as $T_b = \delta\, T_{b,int}$, where $\delta$ is the
\textit{Doppler boosting factor}. Since $T_{b,int}$ is expected to lie below the inverse-Compton limit, our result implies that the source
is indeed relativistically Doppler boosted i.e., $\delta>1$ (see also \cite{antao2016}).

\section{A Bayesian criterion for the resolution limit}
\label{sec:martividal}
\citet{martividal2012} derived the maximum theoretical over-resolution power of an interferometer, dependent 
on the signal-to-noise ratio of the visibility measurements, from theoretical considerations. The minimum resolvable source size 
$\theta_{M}$ is given by
\begin{equation}
\label{eq:martividal}
\theta_M = \beta \left( \frac{l_c}{2(\mathrm{SNR})^2}\right)^{1/4}\ \times\ \mathrm{FWHM}\quad.
\end{equation}
$\beta$ depends on the shape and the intensity profile of the source model and takes values between 0.5 and 1. For source profiles with higher 
intensities at lower scales, the value of $\beta$ is closer to 1. $l_c$ is the log-likelihood value
that corresponds to the critical probability of the null hypothesis, which is taken to be the point source model. It takes the values 3.84 and 
8.81 for a 5 per cent and a 0.3 per cent probability cut-off of the null hypothesis respectively \citep{martividal2012}.
The FWHM is the \textit{full-width at half maximum} of a circular PSF. For an interferometer with an elliptical PSF such as ours, 
the circular FWHM equivalent is given by $\sqrt{ab}$, where $a$ and $b$ are the major and minor axes of the PSF respectively \citep{lobanov2005}. 

The SNR is calculated by computing the ratio between the weighted average of the visibilities and the noise $\sigma = 
\sigma_{\mathrm{vis}}/\sqrt{N}$. 
If $\sigma$ varies between measurements (equation \ref{eq:varyingsigma}), then by \textit{Parseval's theorem}\footnote{Parseval's 
theorem, in this context, ensures that the total power of the noise remains the same 
regardless of whether it is computed in the image domain or in the \textit{uv}-domain.}, the rms noise $\sigma_{\mathrm{rms}}$ in the 
naturally-weighted residual image of the sky (after the source has been subtracted out) can be equated to $\sigma$ as
\begin{equation}
\label{eq:parseval}
\begin{aligned}
\sigma^2 &= \frac{\sigma_{\mathrm{vis}}^2}{N} = \sigma_{\mathrm{rms}}^2\quad,\\
\sigma &= \sigma_{\mathrm{rms}}\quad.
\end{aligned}
\end{equation}

In the Bayesian approach, we set the minimum resolvable size, $\theta_B$, to the size of the source at which the evidence 
for CIRC against PT turns \textit{positive} (Table \ref{tab:evidences}).
To compare $\theta_M$ with $\theta_B$, we simulated a series of observations, each with a compact circular Gaussian source of a different 
size located at the pointing centre of the interferometer. For the station gains and SEFDs, we used the 
\textit{maximum a posteriori} (MAP) estimates\footnote{A MAP estimate is the Bayesian equivalent of the Maximum-Likelihood (ML) estimate and 
corresponds to a mode of the corresponding posterior distribution \citep{sivia2006}. It may be seen as a regularised ML estimate.} obtained 
from analysing the actual data.
We analysed three such sets of simulated data for three different SNR levels, where SNR is calculated according to 
equation (\ref{eq:parseval}), and computed the Bayes factor between CIRC and PT (Table \ref{tab:martividal}).
\begin{table}
\caption{A comparison of the resolution limits denoted by $\theta_M$ and $\theta_B$, obtained using equation (\ref{eq:martividal}) with 
$\beta=1$ and $l_c=8.81$ and by using the Bayesian approach respectively, for the sparse VLBI array described in Table \ref{tab:stations} 
for different SNR levels at 5 GHz. The circular FWHM equivalent of the naturally-weighted elliptical PSF is 
$\sqrt{5.7 \times 2.2} = 3.54$ mas. For higher SNR levels, $\theta_B$ is limited more by the 
gain amplitude calibration than by the theoretical capabilities of the array. The last column gives the maximum brightness temperature one
could measure for a source of 1 Jy, derived from $\theta_B$.}
\begin{center}
\begin{tabular}{ c | c | c | c | c}
\hline
SNR & $\theta_M$ (mas) &  $\theta_B$ (mas) & $\theta_B$/FWHM & $T_b$/$S_{\nu}$ ($10^{12}\, K$/Jy)\\
\hline
150 & 0.42 & 0.45 & 0.13 & 0.763\\
2000 & 0.11 & 0.29 & 0.09 & 1.837\\
5900 & 0.07 & 0.17 & 0.05 & 5.347\\
\hline
\end{tabular}
\end{center}
\label{tab:martividal}
\end{table}

Equation (\ref{eq:martividal}) gives an estimated resolution limit assuming perfect calibration. 
\citet{martividal2012} observe how various factors such as the proportion of long baselines in an array configuration, source 
structure, and biased gains of the antennas providing the long baselines will further limit the resolving power of the interferometer.
The Bayesian approach is sensitive to factors such as source shape, biased instrumental gains, and the associated uncertainties and provides a more 
realistic estimate of the minimum resolvable source size.

\section{Conclusion}
Based on the Bayes factors obtained, we have very strong evidence for a resolved source with slightly elongated structure, in
agreement with what is expected for a partially synchrotron self-absorbed compact jet in a flaring blazar.
By simultaneously estimating source parameters along with the antenna gains and SEFDs, we have also acquired knowledge
of the precision of our estimate of the source size and its correlation with antenna gains. Without SH visibilities, the Bayesian 
evidence indicates that the source must be considered unresolved.

$|g_{\textsc{sh}}|$ varies by about 12 per cent, which is about an order of magnitude worse than the precision in the 
gain amplitude calibration of all but one (NT) European station. The dependence of the shape parameters on $|g_{\textsc{sh}}|$ results in the 
minor axis and the position angle of the Gaussian in model GAU being poorly constrained. 
This also illustrates the necessity of accounting for calibration errors for the stations providing the long baselines, lest we are led astray in our 
attempts to estimate the parameters relevant for the science goals.

\referee{Currently, we are limited only by the performance of the M\textsc{ulti}N\textsc{est} sampler required to compute the evidence.
The RIME can model any time or frequency variation in the source and in the instrumental effects. Within this framework, data from multiple 
spectral bands and epochs can be analysed together, incorporating time and frequency variation in the complex antenna gains. 
Future analyses will benefit from numerical samplers for evidence computation tailored for higher-dimensional parameter spaces.}

The brightness temperature distribution we have derived for J0809+5341 accounts for the uncertainties in the source shape and instrumental gain calibration
and indicates that the source is Doppler boosted and that the intrinsic brightness temperature must be less than what we obtain, consistent
with the literature.

The analysis of synthetic observations of different SNR levels shows that, as the SNR improves (SNR$\gg$100), we are constrained less by the 
theoretical capabilities of the interferometer array and more by the effects of miscalibration of station gains and the discrepancy between 
$\theta_M$ and $\theta_B$ becomes more pronounced.
The constraints derived by \citet{martividal2012} assume perfect calibration. Our approach is more sensitive to uncertainties in source 
shape estimates and their correlation with instrumental gains and provides more realistic estimates that are 2 to 2.5 times the estimates
obtained from equation (\ref{eq:martividal}). Knowledge of $\theta_B$ also enables us to estimate the maximum brightness temperature that one
can measure for a given interferometer configuration.
For sensitive future VLBI arrays such as the SKA-VLBI, it is possible to resolve source structure down to, or even less than, about 5 per cent of the size of the 
naturally weighted restoring beam with high precision, if the gain calibration of the stations providing the longest baselines is precise to within 1 per cent.

\section*{Acknowledgements}

The EVN is a joint facility of independent European, African, Asian, and North American radio astronomy institutes. The data presented in
this publication are derived from the EVN project code RSC02. We are grateful to An Tao and his group for sharing these data
with us and to Ivan Mart\'i-Vidal for his helpful comments. 
We also thank the referee for their comments.
Part of the research leading to these results has received funding from the 
European Commission Seventh Framework Programme (FP/2007-2013) under grant agreement No. 283393 (RadioNet3).

IN and KH acknowledge funding from the MeerKAT High Performance Computing for Radio Astronomy Programme.
JZ and IN acknowledge the South Africa National Research Foundation Square Kilometre Array Project for financial support.
OS is supported by the South African Research Chairs Initiative of the Department of Science and Technology and National Research 
Foundation. Most of the computations leading to these results were carried out on Rhodes University's RATT (Radio Astronomy Techniques \&
Technologies) servers.




\bibliographystyle{mnras}
\bibliography{references} 


\bsp	
\label{lastpage}
\end{document}